\documentclass[a4paper]{jpconf}
\usepackage[english]{babel}
\usepackage{graphicx}
\usepackage{verbatim}
\usepackage{amsfonts}
\usepackage{makeidx}
\usepackage{color}
\usepackage{amsmath}
\usepackage{lineno}
\usepackage{iopams}
\usepackage{epsfig}
\usepackage{hyperref}
\usepackage{cite}

\newcommand{\sqrts}{\sqrt{s}}

\newcommand{\DtoKpi}{{\rm D}^0 \to {\rm K}^-\pi^+}
\newcommand{\DtoKpipi}{{\rm D}^+\to {\rm K}^-\pi^+\pi^+}
\newcommand{\DstartoDpi}{{\rm D}^{*+} \to {\rm D}^0 \pi^+}

\newcommand{\Dzero}{{\rm D^0}}

\newcommand{\Dstar}{{\rm D^{*+}}}

\newcommand{\Dplus}{{\rm D^+}}

\begin{document}
\title{Measurement of the multiplicity dependence of charm production in proton--proton collisions at $\sqrts=7$~TeV with the ALICE experiment at the CERN-LHC}

\author{Aamer Wali Rauf (for the ALICE Collaboration)}

\address{Department of Physics, COMSATS Institute of Information Technology, Islamabad, Pakistan}

\ead{aamer\_wali@comsats.edu.pk}

\begin{abstract}
Potential of the charm quark as a probe to study the Quark-Gluon Plasma (QGP) is best harnessed when its production mechanisms are disentangled from its propagation through the QGP. Proton-proton (pp) collisions help us to study charmed hadron production mechanisms. The measurement of D-meson yields in pp collisions as a function of the multiplicity of produced particles allows one to gain some insight into the processes occurring in the collision at a microscopic level. Here, the preliminary results are presented from this measurement at $\sqrts$ = 7 TeV. The analysis strategy, the applied corrections, and  the determination of the systematic uncertainties are described. The preliminary results are presented and compared with those from a similar, published, measurement of J/$\psi$ production.
\end{abstract}

\section{Introduction}
The study of the D-meson yield dependence on charged particle multiplicity in $\rm pp$ collisions helps to gain better understanding of the charmed hadron production mechanisms. A correlation between the yield of heavy quarks and the total charged particle multiplicity would help to choose between the two possible interpretations, namely an increased hadronic activity in collisions where heavy quarks are produced~\cite{Palaiseau}, and Multi-Parton Interactions (MPI)~\cite{Sjostrand,Bartalini}. Note that the MPI so far have been considered to involve only light quarks and gluons, however the present study may help to determine the quantitative contribution, if any, of the MPI on a harder scale~\cite{Porteboeuf} at the LHC energies. The ALICE collaboration at the LHC has already observed~\cite{JpsiMultPaper} an approximately linear increase of the J/$\psi$ yield with increasing multiplicity for pp collisions at $\sqrts=7$~TeV in the two different rapidity regions: mid-rapidity $(|y| < 0.9)$ and forward-rapidity $(-4.5 < y < -2.4)$. The study is performed in different intervals of D meson transverse momentum with five multiplicity intervals in each $p_{\rm T}$ bin.      

The proceedings are organized as follows. Section 2 describes the ALICE detector and the data sample used in the analysis. In section 3, the multiplicity estimator is discussed. The D-meson analysis strategy along with corrections and systematic uncertainties forms section 4. The results are presented in section 5 followed by concluding remarks in section 6.   

\section{ALICE detector and data sample}
The ALICE detector setup is described in detail in~\cite{aliceJINST}. Results presented here were extracted using the data obtained with the following sub-detectors.

The Inner Tracking System (ITS) comprises six cylindrical layers, two each of silicon pixel (SPD), silicon drift, and silicon strip detectors. They cover the pseudorapidity range $|\eta|<0.9$ for radial distances from the beam axis in the range between 4 and 43 cm. The high segmentation and the positioning of the ITS result in high spatial resolution which allows to reconstruct the decay vertices of charm and beauty hadrons, typically separated by few hundreds of micrometers from the primary vertex. The SPD also provides a collision trigger.
Surrounding the ITS is the cylindrical Time Projection Chamber (TPC)~\cite{TPCcalib} spanning radii from 85 cm to 247 cm. It is the main tracking device that provides track reconstruction with up to 159 three-dimensional space points per track, as well as particle identification through a specific energy loss dE/dx measurement.
The charged particle identification capability of the TPC is supplemented by the cylindrically configured Time-Of-Flight (TOF) detector that occupies radii 370-399 cm from the beam axis. In the current analysis it provides the kaon/pion separation up to $p_{T} \sim$ 2 GeV/${\it c}$. Results from the TOF commissioning with cosmic-ray particles are reported in~\cite{TOFcosmics}. 

The results shown here employ 300 million minimum-bias events collected during the 2010 LHC run with pp collisions at $\sqrts$ = 7 TeV. The minimum-bias trigger required at least one hit in either of the VZERO counters or in the SPD in coincidence with the arrival of proton bunches from both directions. The VZERO detector is composed of two arrays of scintillators covering the full azimuth in the pseudorapidity regions $2.8 < \eta < 5.1$ (VZERO-A) and $-3.7 < \eta < -1.7$ (VZERO-C).    

\section{Multiplicity}
The multiplicity estimator used in the analysis is the number of SPD tracklets for $|\eta|<0.9$. An SPD tracklet is a straight line joining space points in the two layers, and pointing to the reconstructed primary vertex. The measured raw number of tracklets is corrected on an event-by-event basis for the acceptance of the SPD detector, which depends on the position of the primary vertex along the beam axis. Furthermore, the correction accounts for the variation of the SPD acceptance with time during the data taking period, due to the changes in the number of active modules. The D meson yields are extracted in five intervals of multiplicity. As the simulations show linearity between the number of tracklets and the charged particle pseudorapidity density~\cite{JpsiMultPaper}, ${\rm d}N_{ch}/{\rm d}\eta$, the present analysis shows the multiplicity as charged particle pseudorapidity density, normalized to the average pseudorapidity density in pp collisions, that is, ${\rm d}N_{ch}/{\rm d}\eta/<{\rm d}N_{ch}/{\rm d}\eta>$. The definition of the charged particle multiplicity intervals and the corresponding ${\rm d}N_{ch}/{\rm d}\eta$ was done consistently with that of the analysis of J/$\psi$ production as a function of multiplicity~\cite{JpsiMultPaper}. 

\section{D meson analysis} 
\subsection{Raw yield extraction}
$\Dzero$, $\Dstar$, and $\Dplus$ mesons and their anti-particles were reconstructed in the central
 rapidity region from their charged hadronic decay channels $\DtoKpi$ (with branching ratio, BR, 
 of $ (3.88 \pm 0.05)\%$ and decay length $c\tau$= 123 $\mu$m), $\DstartoDpi$ (with BR = $(67.7\pm0.05)\%$) and $\DtoKpipi$ (BR = $(9.13\pm0.19)\%$, $c\tau$= 312 $\mu$m). 
The analysis strategy is based on an invariant mass analysis of fully reconstructed decay topologies displaced from the primary vertex. The statistical significance of the signal over the large combinatorial background was enhanced by means of kinematical and topological selections (see~\cite{ALICEDpp7TeV} for detail), exploiting in particular the separation between the secondary and primary vertex and the pointing of the reconstructed D-meson momentum to the primary vertex. In addition, particle identification of the D-meson decay products based on their specific energy loss and time of flight measurements with the TPC and the TOF, respectively, further suppressed the combinatorial background.

D-meson yields were extracted in different $p_{T}$ ranges, and for a given $p_{T}$ interval in different multiplicity ranges. The $p_{T}$ ranges used in this analysis were [1,2], [2,4], [4,8], [8,12], and [12,20] GeV/${\it c}$ whereas the multiplicity bins in each $p_{T}$ interval were defined as: [1,8], [9,13], [14,19], [20,30], and [31,49]. The raw yields were extracted by fitting the invariant mass distributions with functions composed of a Gaussian for the signal and appropriate function describing the combinatorial background. The position and width of the Gaussian functions in the multiplicity bins were fixed to the values extracted from the multiplicity integrated distribution where the statistical significance of the signal is larger.    

\subsection{Corrections and systematic uncertainties}
The D-meson raw yields were corrected for their kinematical, topological, and PID selection efficiencies as well as for geometrical acceptance $(A \times \epsilon)$. The results shown in the next section are presented in the form of relative yields, $({\rm d^{2}}N^{D}/{\rm d}y{\rm d}p_{T})/<{\rm d^{2}}N^{D}/{\rm d}y{\rm d}p_{T}>$. The yields in each ($p_{T}$, multiplicity) bin were normalized to the minimum-bias yield in each $p_{T}$ bin.  
\begin{linenomath*}
\begin{equation}
\frac{{\rm d^{2}}N^{D}/{\rm d}y{\rm d}p_{T}}{<{\rm d^{2}}N^{D}/{\rm d}y{\rm d}p_{T}>} = \frac{  Y^{\rm RawMultBin}_{\rm D} / ((A \times \epsilon)^{\rm MultBin} \times N_{\rm event}^{\rm mult} ) } 
        { Y^{\rm RawMultInt}_{\rm D} / ( (A \times \epsilon)^{\rm MultInt } \times N_{\rm event}^{\rm MultInt } )}. 
        \label{eq:CorrYields}
\end{equation}
\end{linenomath*}

Here $\rm MultBin$ means a particular multiplicity bin in a $p_{T}$ bin whereas $\rm MultInt$ denotes a $p_{T}$ bin with all the multiplicity bins integrated. The number of events used for the normalization of the D-meson yield is corrected for the fraction of inelastic events not seen by the minimum-bias trigger~\cite{MBInel}. Simulations were used to determine the acceptance and efficiency correction factor ($A\times\epsilon$). 

Sources of systematic uncertainties include yield extraction, particle identification (PID), topological cuts, MonteCarlo (MC) simulations, and feed-down from B-meson decays.
The yield extraction systematics was estimated by determining yields varying the background fit function and the fit range in mass, by leaving the position and width of the Gaussian as free parameters, and using an alternative technique based on counting the entries in the histogram after background subtraction. Note that the uncertainties in $Y^{\rm RawMultBin}_{\rm D}$ and $Y^{\rm RawMultInt}_{\rm D}$ were assumed to be uncorrelated, and, therefore, added in quadrature. It was estimated to be (1-10\%) depending on the meson species, and the ($p_{T}$, multiplicity) bin. The corrected yields as given in Eq.~\ref{eq:CorrYields} were determined with and without applying the PID selections. The ratio of the yields with and without PID was found to be compatible with unity in all the ($p_{T}$, multiplicity) bins. Consequently, no systematic uncertainty related to the PID selection was considered. The systematic uncertainty caused by the given topological selections of the D mesons were estimated by studying the relative variations of the corrected results (Eq.~\ref{eq:CorrYields}) with diverse sets of cuts. Such variations amounted to (5-10\%) systematic errors depending on the ($p_{T}$, multiplicity) bin. This is due to the fact that the efficiency corrections, determined from Monte Carlo simulations, depend on the agreement between real and simulated data as far as signal $p_{T}$ shape, distributions of primary particles and detector descriptions are concerned. The systematic uncertainty related to the multiplicity dependence of the D-meson reconstruction and selection efficiency was estimated from the difference in the corrected yield obtained using the efficiencies extracted from two different MC productions, differing for the multiplicity distribution of primary particles. The resulting systematic uncertainty was 10\%. Monte Carlo simulations were also used to verify the linear dependence of the number of reconstructed SPD tracklets and the generated density of primary particles ${\rm d}N_{ch}/{\rm d}\eta$, which was found to be linear within 5\%. 
The corrected prompt yields given by Eq.~\ref{eq:CorrYields} assume that the relative contribution of B decays to the D yields in different multiplicity bins is constant and equal to that of the multiplicity integrated sample. This causes the correction factor accounting for the fraction of prompt D meson yield to cancel in the ratio between the yield in a given multiplicity bin and that in the multiplicity integrated sample. That however might not be the case in reality.
A systematic uncertainty due to the possible different multiplicity dependence of charm and beauty production was included. It was estimated using the fraction of prompt D mesons computed using FONLL predictions, as described in ~\cite{ALICEDpp7TeV} and an hypothesis on the multiplicity dependence of B/D hadron production. 

\section{Results}  
\begin{figure}[h]
\begin{minipage}{18pc}
\includegraphics[width=18pc]{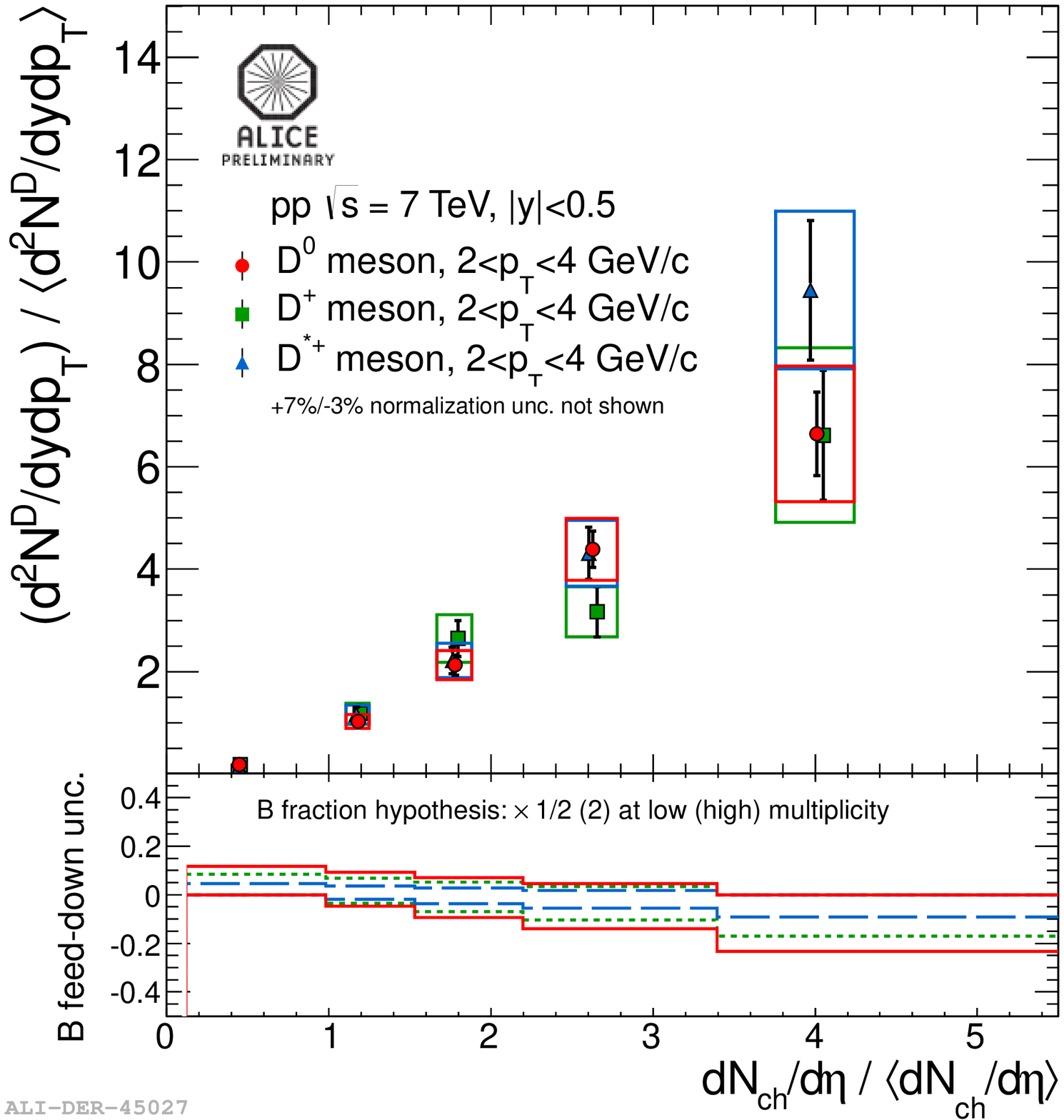}
\label{fig:Dlin}
\end{minipage}\hspace{2pc}
\begin{minipage}{18pc}
\includegraphics[width=18pc]{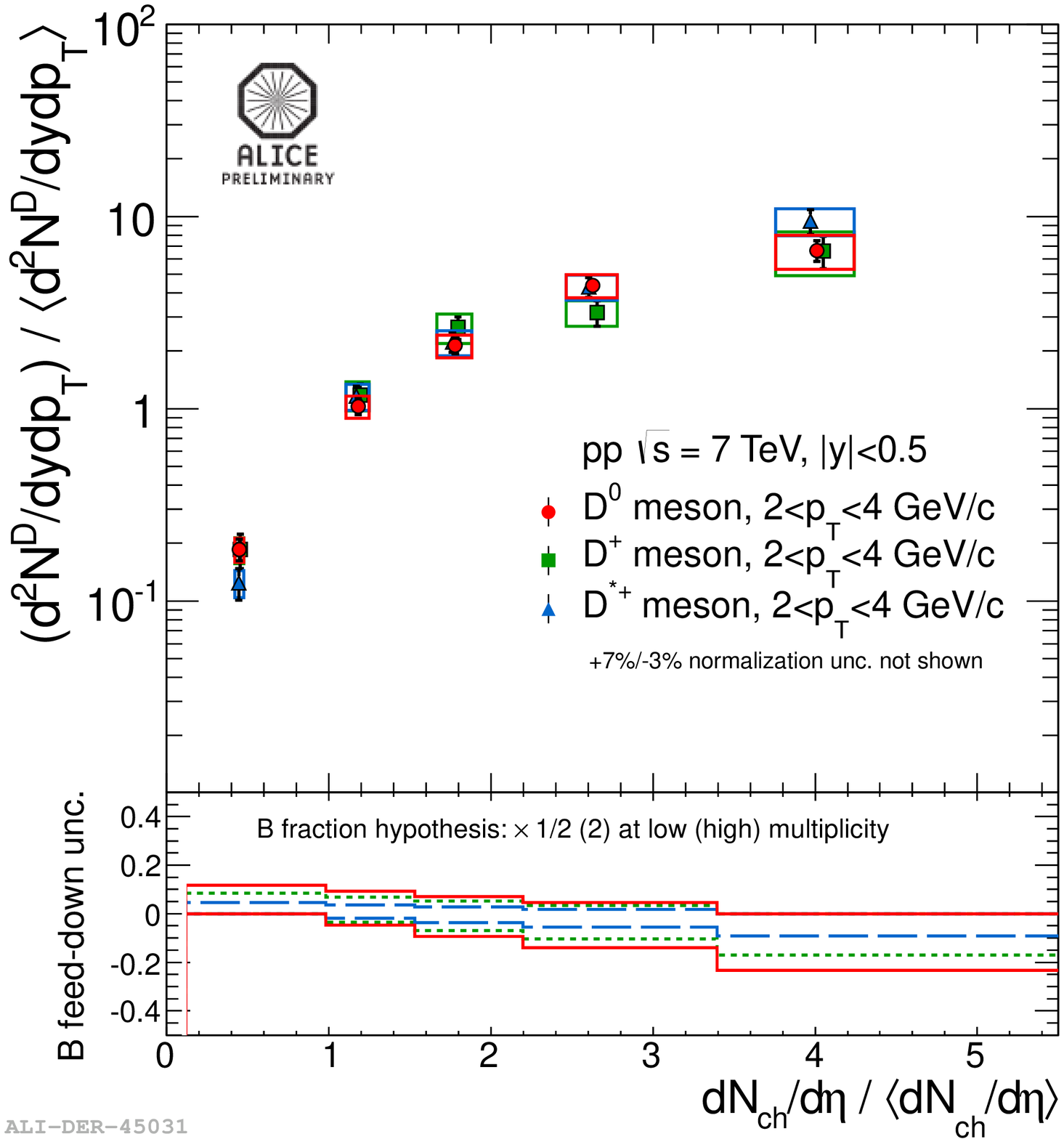}
\label{fig:Dlog}
\end{minipage} 
\caption{\label{fig:Dyield}Relative ${\rm D}^{0}$, ${\rm D}^{+}$ and ${\rm D}^{*+}$ yields in $2<p_{T}<4$ GeV/${\it c}$ as a function of the multiplicity of charged particles produced in $\rm pp$ collision at $\sqrts$ = 7 TeV.}
\end{figure}

\begin{figure}[h]
\begin{minipage}{18pc}
\includegraphics[width=18pc]{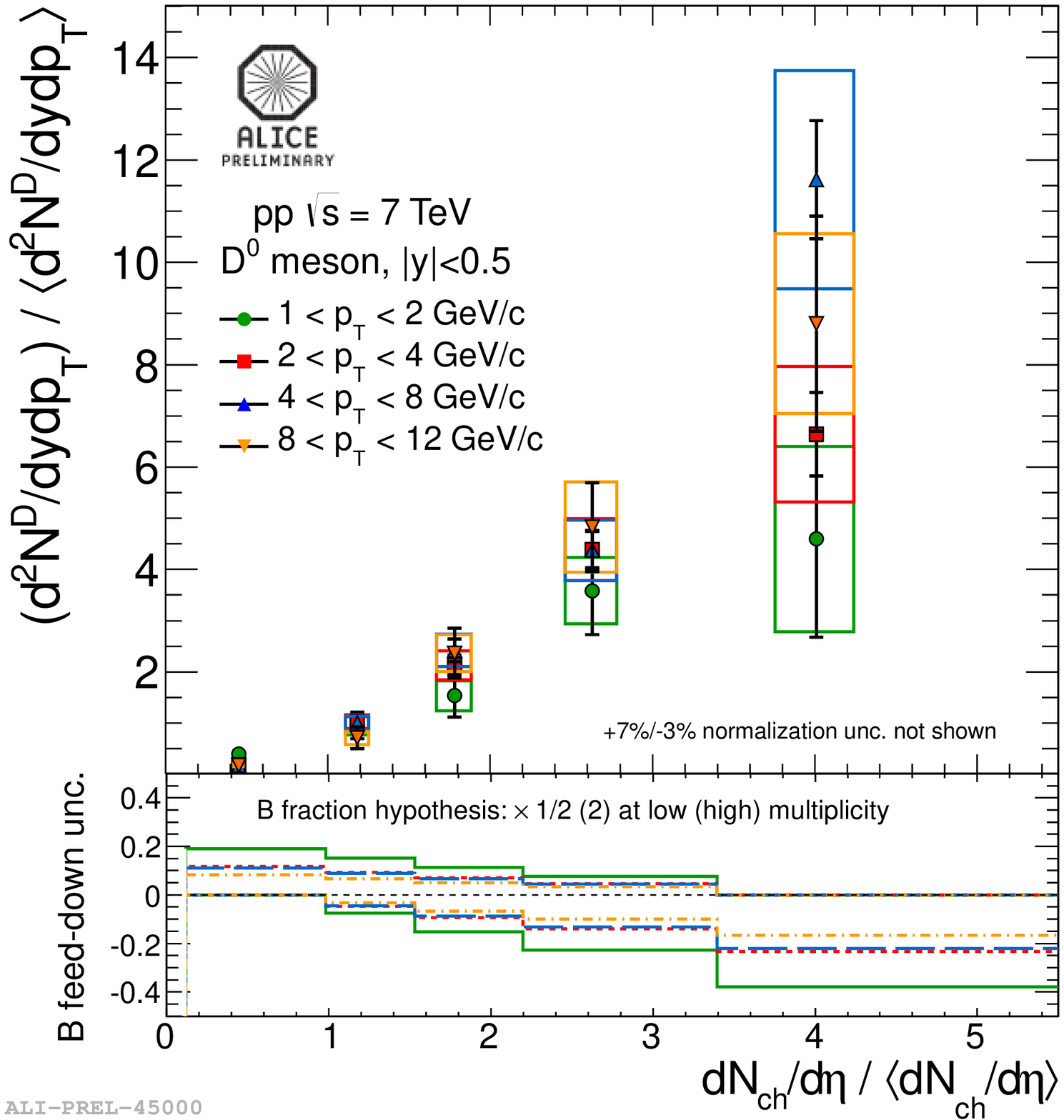}
\label{fig:Dstarlin}
\end{minipage}\hspace{2pc}
\begin{minipage}{18pc}
\includegraphics[width=18pc]{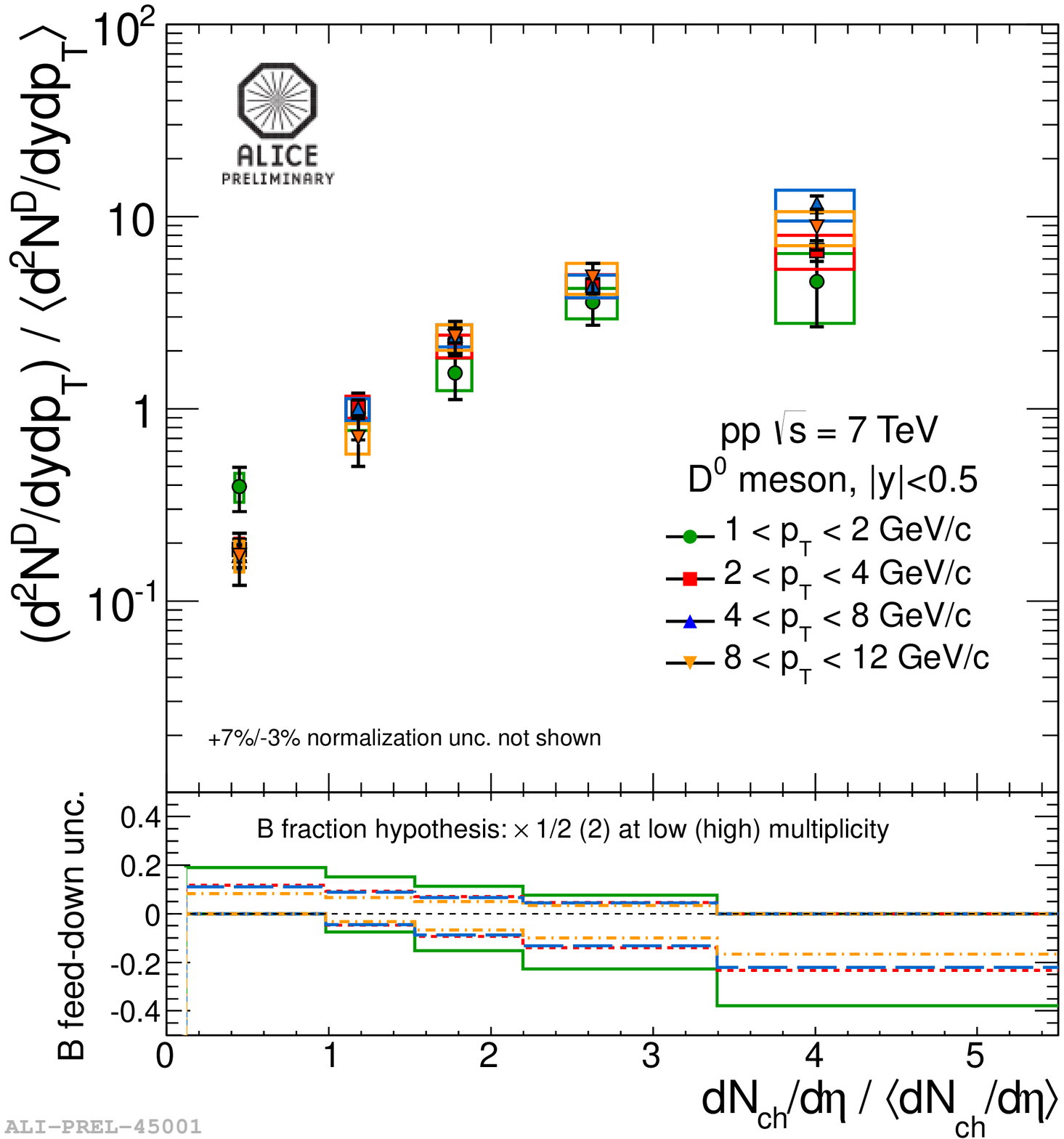}
\label{fig:Dstarlog}
\end{minipage}
\caption{\label{fig:Dstar}Relative ${\rm D}^{0}$ yields in 4 different $p_{T}$ intervals as a function of the multiplicity of charged particles produced in $\rm pp$ collision at $\sqrts$ = 7 TeV.}
\end{figure}

The relative yields for the three D meson species studied as a function of relative charged particle density, ${\rm d}N_{ch}/{\rm d}\eta/<{\rm d}N_{ch}/{\rm d}\eta>$, are shown in Fig.~\ref{fig:Dyield} for the transverse momentum interval $2<p_{T}<4$ GeV/${\it c}$. The two panels in this and the following figures differ only in the y-scale being linear and logarithmic respectively. The horizontal size of the boxes encompasses systematic uncertainty in ${\rm d}N_{ch}/{\rm d}\eta/<{\rm d}N_{ch}/{\rm d}\eta>$. The vertical size of the boxes reflects all the systematic uncertainties but the feed-down in the relative D meson yields. The feed-down systematic uncertainties are shown in the lower panel of each figure. These uncertainties were determined by considering that the relative fraction of the B/D hadrons could vary with the charged particle density by a factor of 0.5 to 2 in all multiplicity bins. The trends observed for the three meson species agree within statistical and systematic uncertainties. Furthermore, an almost linear increase of the yield with multiplicity is observed, within uncertainties, for all the considered $p_{T}$ intervals. Note that, as demonstrated in Fig.~\ref{fig:Dstar}, in the $p_{T}$ bin [1,2] the ${\rm D}^{0}$-meson at low multiplicity does not follow this general trend. 

\begin{figure}[h]
\begin{minipage}{18pc}
\includegraphics[width=18pc]{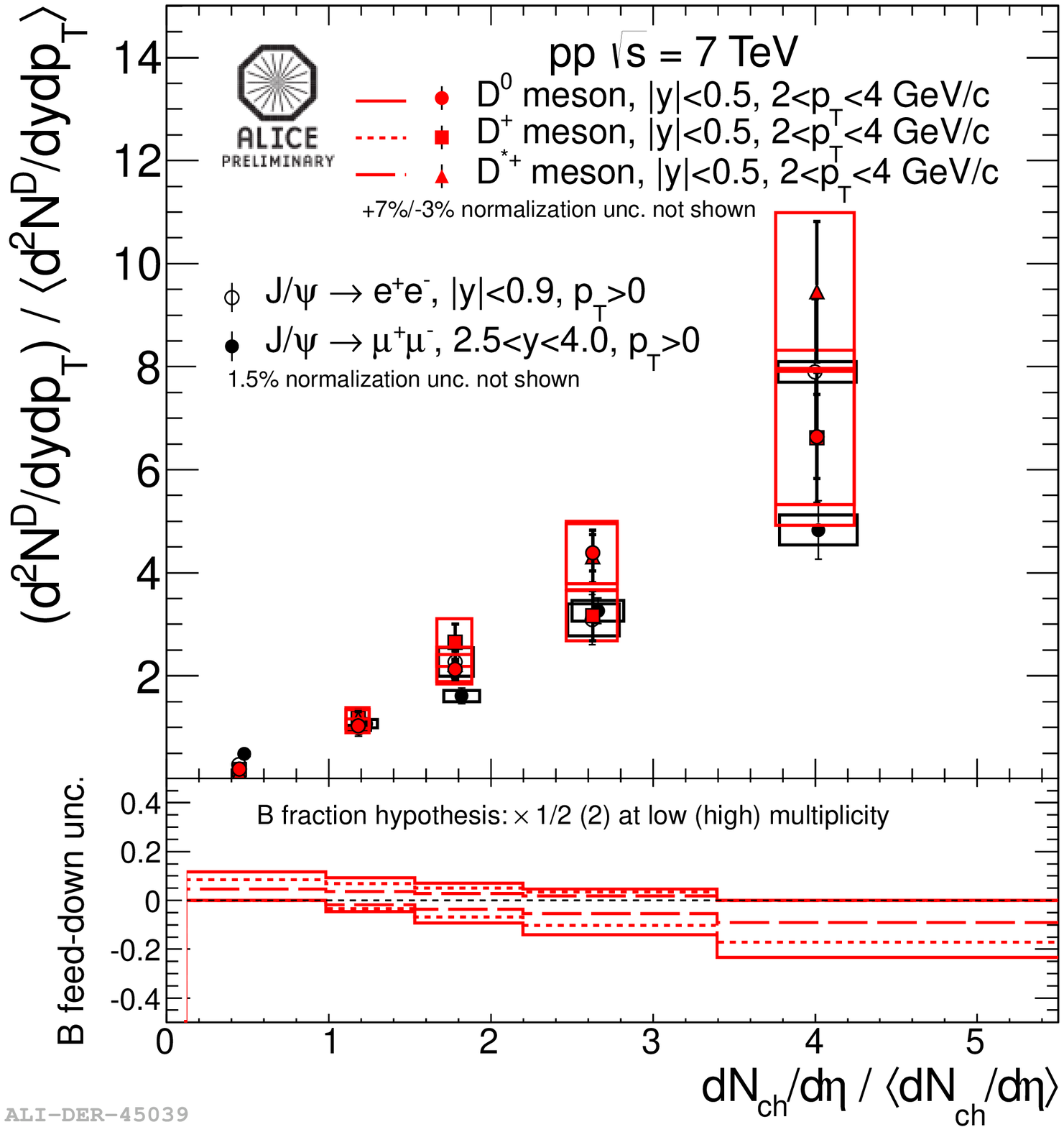}
\label{fig:DJpsilin}
\end{minipage}\hspace{2pc}
\begin{minipage}{18pc}
\includegraphics[width=18pc]{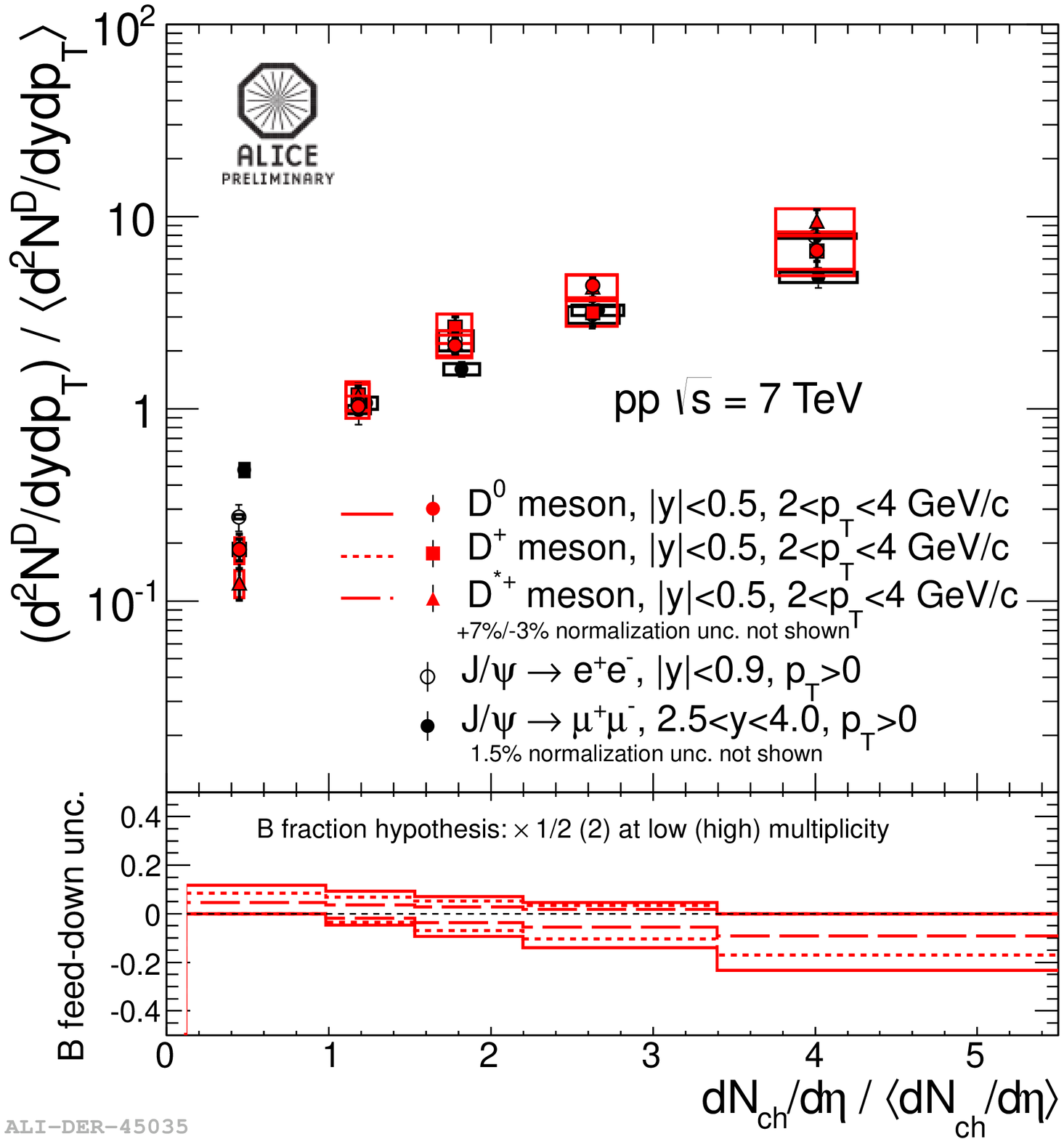}
\label{fig:DJpsilog}
\end{minipage}
\caption{\label{fig:DJpsi}Relative ${\rm D}^{0}$, ${\rm D}^{+}$ and ${\rm D}^{*+}$ yields in $2<p_{T}<4$ GeV/${\it c}$ as a function of the multiplicity compared to J/$\psi$ results at central and forward rapidity in $\rm pp$ collision at $\sqrts$ = 7 TeV.}
\end{figure}

\begin{figure}[h]
\begin{minipage}{18pc}
\includegraphics[width=18pc]{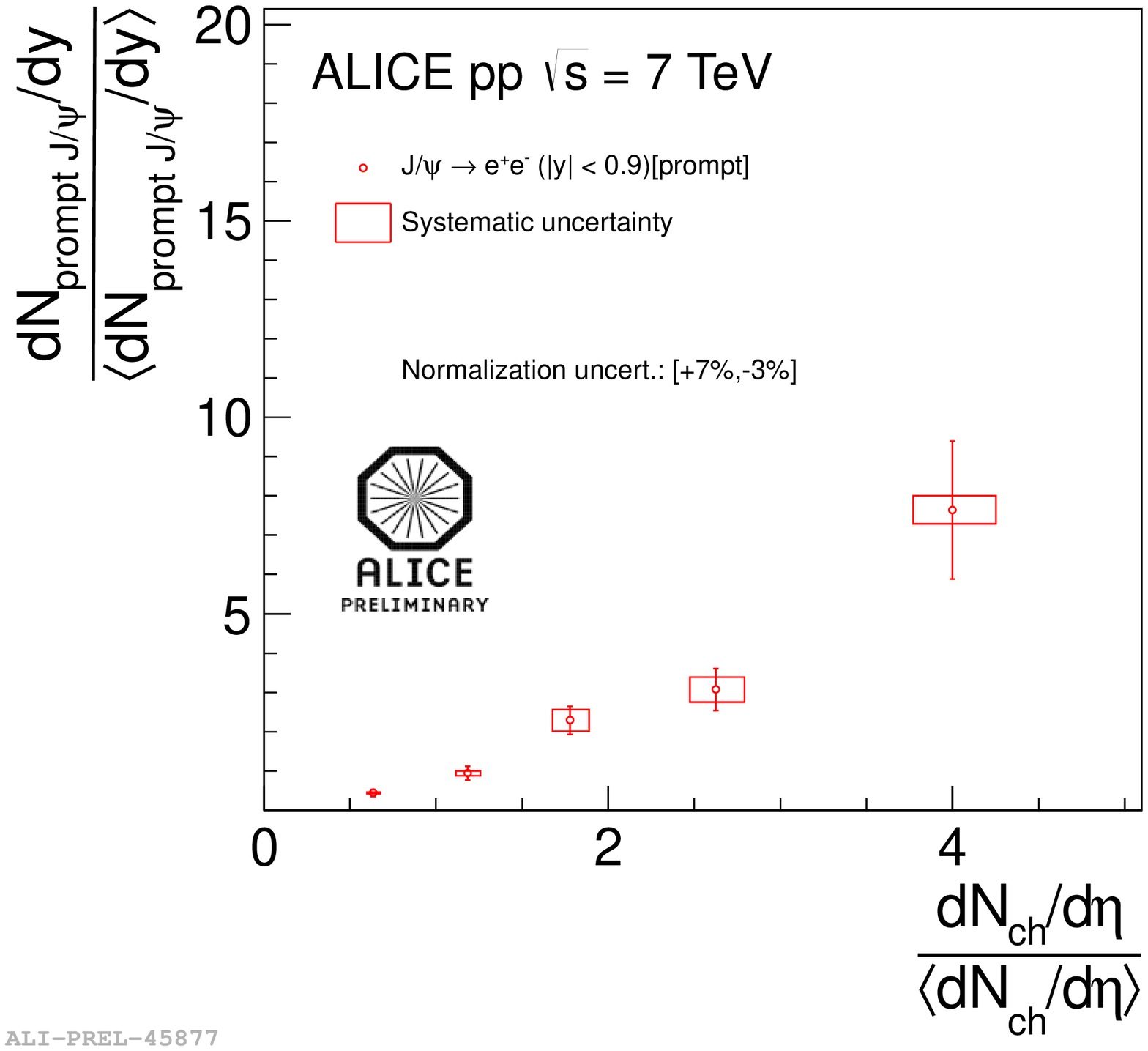}
\label{fig:JpsiPrompt}
\end{minipage}\hspace{2pc}
\begin{minipage}{18pc}
\includegraphics[width=18pc]{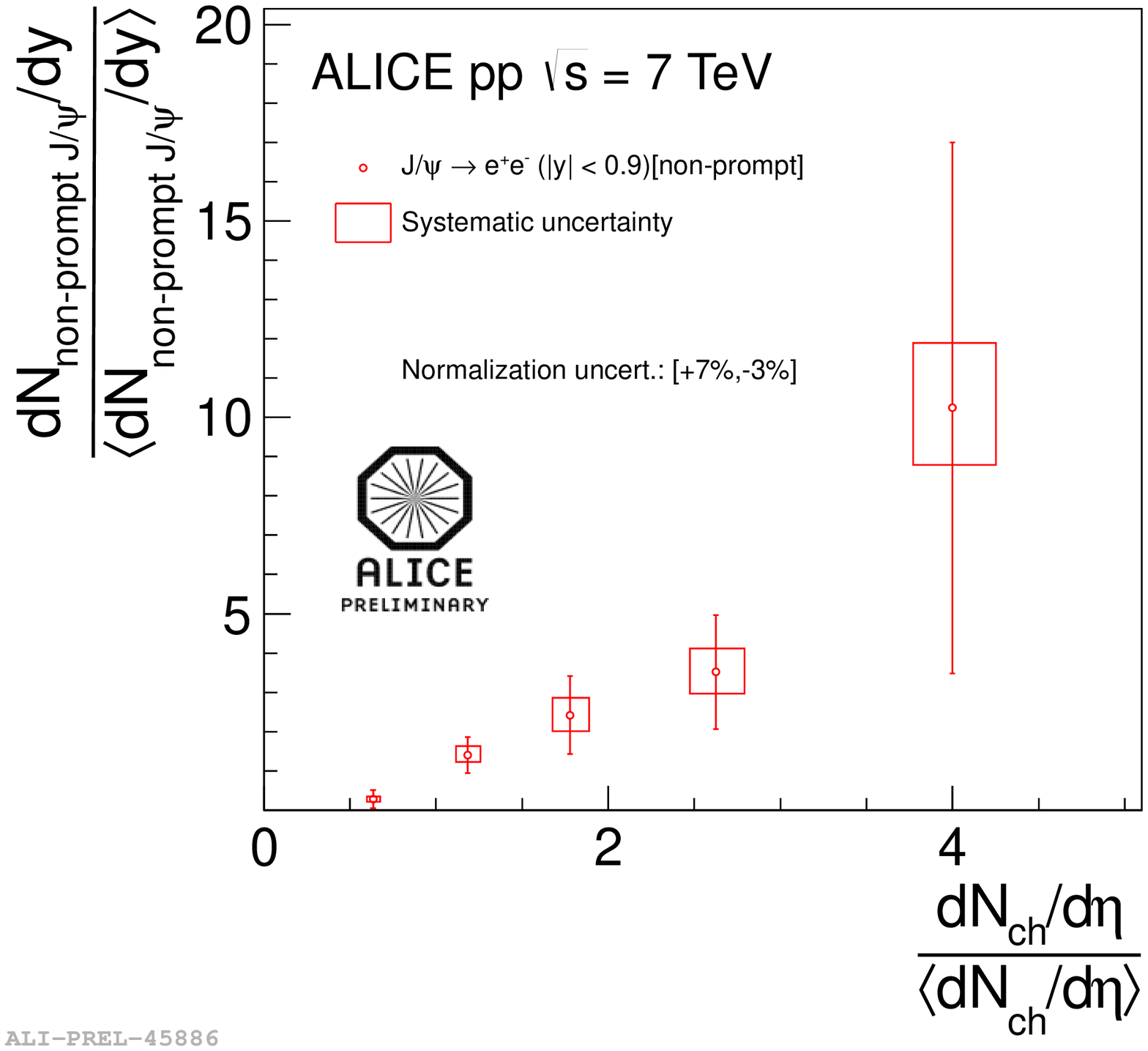}
\label{fig:JpsiNonPrompt}
\end{minipage}
\caption{\label{fig:Jpsi}Prompt and non-prompt relative J/$\psi$ yields versus relative multiplicity in $\rm pp$ collision at $\sqrts$ = 7 TeV.}
\end{figure}

Fig.~\ref{fig:DJpsi} shows the comparison of D-meson yields with the inclusive J/$\psi$ yield at $\sqrts$ = 7 TeV. The J/$\psi$ results are $p_{T}$ integrated over two rapidity intervals $|y|<0.9$ and $2.5<y<4.0$ whereas the D-meson rapidity and $p_{T}$ intervals are $|y|<0.5$ and 2-4 GeV/${\it c}$ respectively. The results for open and hidden charm mesons agree within uncertainties. It is interesting to note in Fig.~\ref{fig:Jpsi} that both the prompt (direct J/$\psi$ production from charm) and the non-prompt (the feed-down contribution to J/$\psi$ from B mesons) components show approximately similar linear behaviour with multiplicity.

\section{Conclusion}
We have studied the dependence of $D^{0}$, $D^{+}$ and $D^{*+}$ yields on multiplicity for five D meson $p_{T}$ intervals, namely [1,2], [2,4], [4,8], [8,12], and [12,20] GeV/${\it c}$. The relative yields are found to be consistent with each other for the $p_{T}$ intervals considered in the current analysis. An approximately linear increase of D-meson yields with charged  particle multiplicity is observed. This approximately linear trend is quantitatively consistent for the three D meson species and the five considered $p_{T}$ intervals. We also compared the current results in the $p_{T}$ = 2-4 GeV/${\it c}$ bin with the already published J/$\psi$ measurements~\cite{JpsiMultPaper}.  It should be noted that the D meson measurements were done in the central rapidity interval whereas the J/$\psi$ measurements were obtained in both the central and the forward rapidity region. The comparison demonstrates that both the open and hidden charm mesons follow a similar trend as a function of multiplicity.

\end{document}